\documentclass[12pt]{article}
\usepackage{epsfig}
\def\be{\begin{equation}}
\def\ee{\end{equation}}
\def\bea{\begin{eqnarray}}
\def\eea{\end{eqnarray}}
\usepackage{graphicx}% Include figure files

\catcode`\@=11
\def\lsim{\mathrel{\mathpalette\@versim<}}
\def\gsim{\mathrel{\mathpalette\@versim>}}
\def\@versim#1#2{\vcenter{\offinterlineskip
\ialign{$\m@th#1\hfil##\hfil$\crcr#2\crcr\sim\crcr } }}
\catcode`\@=12

\catcode`@=12
\begin{document}
\thispagestyle{empty}
\begin{flushright}
UCRHEP-T555\\
July 2015\
\end{flushright}
\vspace{0.6in}
\begin{center}
{\LARGE \bf Stable or Unstable Light Dark Matter\\}
\vspace{1.0in}
{\bf Ernest Ma$^1$, M. V. N. Murthy$^2$, and 
G. Rajasekaran$^{2,3}$\\}
\vspace{0.5in}
{\sl $^1$Department of Physics and Astronomy,\\}
%\vspace{0.1in}
{\sl University of California, 
Riverside, California 92521, USA\\}
\vspace{0.2in}
{\sl $^2$The Institute of Mathematical Sciences, Chennai 600 113, India\\}
\vspace{0.2in}
{\sl $^3$Chennai Mathematical Institute, Siruseri 600 103, India}
\end{center}
\vspace{1.0in}

\begin{abstract}\
We consider the case of light dark matter ($\sim 10$ GeV).  We discuss a 
simple $Z_2$ model of scalar self-interacting stable dark matter, as well 
as a related model of unstable long-lived dark matter which can explain 
the anomalous Kolar events observed decades ago.
\end{abstract}

\newpage
\baselineskip 24pt

Searches for the dark matter (DM) of the Universe have so far not yielded 
unambiguous results.  Whereas experimental efforts have concentrated 
on the 100 GeV range of DM masses, there are hints that it may actually 
be much lighter, say $\sim 10$ GeV.  One intensely studied scenario at 
present~\cite{l14} comes from the observation of gamma rays originating 
in the galactic center, which may be due to the annihilation of dark matter 
of mass $\sim 30$ GeV to $b \bar{b}$ or of mass $\sim 10$ GeV to 
$\tau^- \tau^+$.  There is also another much less known hint coming from  
the anomalous Kolar events~\cite{krishna1, krishna2} recorded in underground 
detectors in the 70's and 80's, which have recently been 
interpreted~\cite{mr14} as being due to the decays of DM with mass in 
the range of 5-10 GeV. 

In this paper we consider the case of light dark matter ($\sim 10$ GeV) 
which is stable or unstable.  We discuss two minimal examples.  The first 
involves a complex scalar $\zeta = (\rho + i \sigma)/\sqrt{2}$, such that 
$\sigma$ is stable DM and $\rho$ is a light mediator for the self 
interactions of $\sigma$.  The second has additional singlet charged fermion 
$E$ and singlet neutral scalar $\eta$ such that $\sigma$ becomes unstable 
and decays dominantly to $\mu^- \mu^+$ in one loop with the correct lifetime 
to explain the anomalous Kolar events.

(I) We consider first the model of a complex neutral scalar $\zeta$ 
interacting with the standard model (SM) Higgs doublet $\Phi = 
(\phi^+,\phi^0)$.  Let the scalar potential be given by
\begin{equation}
V = \mu_1^2 \Phi^\dagger \Phi + \mu_2^2 \zeta^* \zeta + {1 \over 2} \lambda_1 
(\Phi^\dagger \Phi)^2 + {1 \over 2} \lambda_2 (\zeta^* \zeta)^2 + \lambda_3 
(\Phi^\dagger \Phi) (\zeta^* \zeta),
\end{equation}
so that $\zeta$ is protected by a $U(1)_D$ symmetry.  If $U(1)_D$ is unbroken, 
$\zeta$ is a possible DM candidate.  However, its mass $\mu_2$ and coupling 
$\lambda_3$ are constrained by three conditions: (1) its thermal relic 
abundance, (2) the upper limit on its elastic scattering cross section 
off nuclei in direct-search experiments, and (3) the invisible decay 
$h \to \zeta^* \zeta$ if $m_\zeta < m_h/2$ where $h$ is the one Higgs boson 
of the SM, identified as the 125 GeV particle discovered~\cite{atlas12,cms12} 
at the Large Hadron Collider (LHC) in 2012.  Note that these are the same 
conditions if $U(1)_D$ is replaced by $Z_2$ and $\zeta$ by a real scalar 
singlet~\cite{sz85}.  As such, for $m_\zeta = 10$ GeV, it is definitely ruled 
out~\cite{fpu15}.

We now add the term
\begin{equation}
V' = {1 \over 2} \mu_3^2 \zeta \zeta + H.c.
\end{equation}
Without loss of generality, we choose $\mu_3^2$ to be real, and define 
\begin{equation}
\zeta = {\rho + i \sigma \over \sqrt{2}}.
\end{equation}
We then have
\begin{equation}
V' = {1 \over 2} \mu_3^2 (\rho^2 - \sigma^2).
\end{equation}
This means that $U(1)_D$ is softly broken to $Z_2 \times Z_2$.  This scenario 
does not differ qualitatively from the unbroken one and is also ruled out.

We now allow the $Z_2$ symmetry associated with $\rho$ be broken spontaneously 
with $\langle \rho \rangle = u$ and redefine $\rho$ as $u + \rho$.  Together 
with replacing $\Phi$ with $(0, (v + h)/\sqrt{2})$, we obtain the 
minimization conditions for $V + V'$:
\begin{eqnarray}
&& \mu_1^2 + {1 \over 2} \lambda_1 v^2 + {1 \over 2} \lambda_3 u^2 = 0, \\
&& \mu_2^2 + \mu_3^2 + {1 \over 2} \lambda_2 u^2 + {1 \over 2} \lambda_3 v^2 
= 0.
\end{eqnarray}
As a result, after dropping a constant term,
\begin{eqnarray}
V + V' &=& {1 \over 2} \lambda_1 v^2 h^2 + {1 \over 2} \lambda_2 u^2 \rho^2 
+ \lambda_3 v u h \rho - \mu_3^2 \sigma^2 
\nonumber \\ 
&+& {1 \over 2} \lambda_1 v h^3 + {1 \over 2} \lambda_2 u \rho 
(\rho^2 + \sigma^2) + {1 \over 2} \lambda_3 v h (\rho^2 + \sigma^2) \nonumber \\ 
&+& {1 \over 8} \lambda_1 h^4 + {1 \over 8} \lambda_2 (\rho^2 + \sigma^2)^2 
+ {1 \over 4} \lambda_3 h^2 (\rho^2 + \sigma^2). 
\end{eqnarray}
Hence $m_\sigma^2 = -2\mu_3^2$ may be chosen to be 10 GeV 
and be DM because it is still protected by an unbroken $Z_2$ symmetry. 
Its coupling to $h$, i.e. $\lambda_3$, may be chosen to be very small 
to satisfy the direct-search bound.  As for its relic density, we now 
have another annihilation channel to consider, i.e. $\sigma \sigma \to 
\rho \rho$, as shown in Fig.~1.
\begin{figure}[htb]
\vspace*{-3cm}
\hspace*{-3cm}
\includegraphics[scale=1.0]{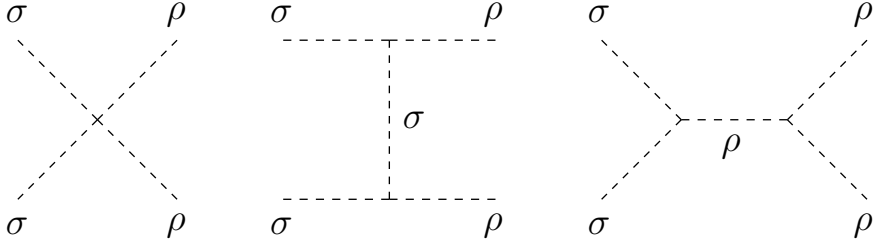}
\vspace*{-21.5cm}
\caption{$\sigma \sigma$ annihilation to $\rho \rho$ final states.}
\end{figure}
The mass-squared matrix spanning $h$ and $\rho$ is given by
\begin{equation}
{\cal M}_{h \rho}^2 = \pmatrix{ \lambda_1 v^2 & \lambda_3 v u \cr \lambda_3 v u 
& \lambda_2 u^2}.
\end{equation}
We look for a solution where $h - \rho$ mixing is small, with two mass 
eigenvalues such that one is 125 GeV and the other 10 MeV. The latter 
may then act as a suitable scalar mediator for $\sigma$ as self-interacting 
dark matter~\cite{s14}.  Since $\lambda_3$ is already assumed to be small, 
the mixing $(\lambda_3/\lambda_1)(u/v)$ will not be significant.  Hence  
$\rho$ is essentially a mass eigenstate with $m_\rho^2 \simeq \lambda_2 u^2$ 
and the amplitude of Fig.~1 is given by 
\begin{equation}
{\cal A} = \lambda_2 - {\lambda_2^2 u^2 \over 4 m_\sigma^2} = \lambda_2 
\left( 1 - {m_\rho^2 \over 4 m_\sigma^2} \right),
\end{equation}
and the annihilation cross section $\times$ relative velocity is
\begin{equation}
\sigma \times v_{rel} = {{\cal A}^2 \over 64 \pi m_\sigma^2}.
\end{equation}
Setting this to the optimal value~\cite{sdb12} 
$2.2 \times 10^{-26}~{\rm cm}^3~{\rm s}^{-1}$ for the correct dark-matter 
relic density of the Universe, we find for $m_\sigma = 10$ GeV and 
$m_\rho = 10$ MeV,
\begin{equation}
\lambda_2 = 0.006, ~~~ u = 0.13~{\rm GeV}.
\end{equation}
Once produced, $\rho$ achieves thermal equilibrium with the SM particles 
through its mixing with $h$ and decays mainly to $e^- e^+$ pairs.

To explain~\cite{mr14} the anomalous Kolar events, $\sigma$ must decay 
with a lifetime slightly greater than the age of the Universe.  Its decay 
product should also be mostly muons.  To make $\sigma$ unstable, 
one option is to add the trilinear term
\begin{equation}
V'' = \mu_4 \Phi^\dagger \Phi \zeta + H.c.
\end{equation}
This induces $h - \sigma$ mixing given by $\sqrt{2} Im(\mu_4)/\lambda_1 v$, 
which is then arbitrarily chosen to be extremely small ($\sim 10^{-19}$) 
so that the $\sigma$ decay lifetime is of order the age of the Universe.  
Even so, its decay must still go through $h$ and thus mainly into 
$\tau^- \tau^+$.   The branching fraction of $\tau$ to $\mu$ diminishes 
the number of observable events to only a few percent, which is not a 
desirable scenario for a robust explanation of the Kolar events, but 
cannot be ruled out with the meager data at hand. 

(II) To obtain a naturally suppressed lifetime and a dominant decay to muons, 
we now consider the following addition to our $U(1)_D$ model.  We add two 
new fields: negatively charged fermion $E_{L,R}$ and neutral complex 
scalar $\eta$, each transforming as $1/2$ under $U(1)_D$.   As a result, 
we have the possible new interactions:
\begin{eqnarray}
{\cal L}' &=& f \bar{E}_L \mu_R \eta + \mu_5 \zeta^* \eta \eta + H.c. 
\nonumber \\ 
&=& {f \over \sqrt{2}} \bar{E}_L \mu_R (\eta_R + i \eta_I) + 
{f \over \sqrt{2}} \bar{\mu}_R E_L (\eta_R - i \eta_I) \nonumber \\ 
&+& {\mu_5 u \over 2 \sqrt{2}} (\eta_R^2 - \eta_I^2) + 
{\mu_5 \over 2 \sqrt{2}} \rho (\eta_R^2 - \eta_I^2) + 
{\mu_5 \over \sqrt{2}} \sigma \eta_R \eta_I,
\end{eqnarray}
where $\eta = (\eta_R + i \eta_I)/\sqrt{2}$ and $\mu_5$ has been chosen real 
by absorbing the arbitrary phase of $\eta$.  This implies 
\begin{equation}
m_R^2 - m_I^2 = {\mu_5 u \over \sqrt{2}},
\end{equation}
and $\sigma$ decays into $\mu^- \mu^+$ as shown in Fig.~2.
\begin{figure}[htb]
\vspace*{-3cm}
\hspace*{-3cm}
\includegraphics[scale=1.0]{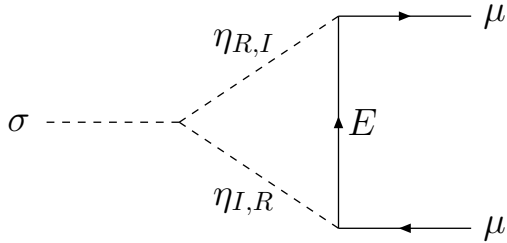}
\vspace*{-21.5cm}
\caption{One-loop $\sigma$ decay to $\mu^- \mu^+$.}
\end{figure}
The effective coupling of $\sigma \bar{\mu} \gamma_5 \mu$ is then 
given by
\begin{equation}
f_{eff} = {f^2 \mu_5^2 u m_\mu \over 
128 \pi^2 m_E^4} \left[ {-2-5 x + x^2 \over 6x(1-x)^3} - {\ln x \over (1-x)^4} 
\right],
\end{equation}
where $x = m_\eta^2/m_E^2$.  Note that $f_{eff}$ is highly suppressed for 
large $m_E$ and would be zero if $m_R = m_I$, i.e. $u=0$.  The decay width 
of $\sigma$ is
\begin{equation}
\Gamma_\sigma = {m_\sigma f^2_{eff}\over 8 \pi}.
\end{equation}
For $\tau_\sigma = 1/\Gamma_\sigma$ to be the lifetime of the Universe, 
i.e. 13.82 billion years, we need 
\begin{equation}
f_{eff} \simeq 2 \times 10^{-21}.
\end{equation}
In the limit $x=1$, the bracket has the value $-1/12$.  Consider for example 
$m_E = 1$ TeV, $u = 0.13$ GeV, $\mu_5 = 1$ GeV, then the above value of 
$f_{eff}$ is achieved with $f = 0.044$. 

So far we have assumed that only $\mu$ couples to $E$.  It is clear that 
$e$ or $\tau$ may do so as well.  Note also that in this model, there is 
still an exactly conserved $Z_2$ symmetry, under 
which $\eta_R, \eta_I$ and $E$ are odd, with all other particles even. 
Suppose $\eta_R$ is the lightest, then its relic density is determined 
by its interaction with the SM Higgs boson $h$ through the 
$\lambda_4 (\eta^* \eta) (\Phi^\dagger \Phi)$ term.  We assume that 
$\lambda_4$ is large enough, so that the $\eta_R$ relic density is 
negligible.  This then opens up the possibility for the unstable $\sigma$ 
to be a light dark-matter candidate which also explains the anomalous 
Kolar events.

To conclude, we have discussed the case of stable or unstable light 
($\sim 10$ GeV) dark matter based on an $U(1)_D$ symmetry which is 
consistent with present constraints.  The simplest model where $U(1)_D$ 
breaks to exact $Z_2$ with a complex scalar singlet allows stable 
self-interacting dark matter.  It may also allow unstable dark matter, 
which decays into $\tau^- \tau^+$ through an extremely small mixing 
($\sim 10^{-19}$) with the Higgs boson.  An alternative is to add one 
heavy charged fermion singlet and one heavy neutral scalar singlet (both 
$\sim 1$ TeV) transforming under $U(1)_D$ such that the dark matter may 
decay in one loop dominantly into $\mu^- \mu^+$ with a highly suppressed 
rate, resulting in a lifetime just greater than the age of the Universe 
to explain~\cite{mr14} the 
anomalous Kolar events~\cite{krishna1,krishna2} observed decades ago.

\medskip

\noindent \underline{Acknowledgment}~:~The work of EM is supported in 
part by the U.~S.~Department of Energy under Grant 
No.~DE-SC0008541.

\end{document}